\documentclass[prl, twocolumn, showpacs, superscriptaddress]{revtex4}
\pdfoutput=1 
\usepackage{graphicx}
\begin{document}

\title{Non-local Conductance Modulation by Molecules:\\  
STM of Substituted Styrene Heterostructures on
H-Terminated Si(100)}

\author{Paul G. Piva}

\altaffiliation{Present Address: Institute for National Measurement Standards, 
NRC,
Ottawa, Ontario, Canada K1A 0R6.}

\author{Robert A. Wolkow}

\altaffiliation{CIFAR Fellow, Nanoelectronics Program.}

\affiliation{National Institute for Nanotechnology, National Research Council of Canada, Edmonton, Alberta T6G 2V4, Canada
and Department of Physics, University of Alberta, Edmonton, Alberta T6G 2J1, Canada
}

\author{George Kirczenow}

\altaffiliation{CIFAR Fellow, Nanoelectronics Program.}

\affiliation{Department of Physics, Simon Fraser
University, Burnaby, British Columbia, Canada V5A 1S6}

\date{\today}


\begin{abstract}

One-dimensional organic heterostructures consisting of contiguous lines of CF$_3$- and OCH$_3$-substituted
styrene molecules on silicon are studied by scanning tunneling microscopy and
{\em ab initio} simulation. Dipole fields of OCH$_3$-styrene molecules are found to enhance conduction through
molecules near OCH$_3$-styrene/CF$_3$-styrene heterojunctions. Those of CF$_3$-styrene depress
transport through the nearby silicon. Thus choice of substituents and their attachment site on host molecules
provide a means of differentially tuning molecule and substrate transport at the molecular scale.     

\end{abstract}
\maketitle

Better understanding and control of molecule-surface interactions
are key to furthering advances in catalysis research, thin film
deposition and processing, chemical sensing, and molecular
electronics.  The scanning tunneling microscope remains an
invaluable tool for studying molecule-surface interactions at the
molecular scale.  Its ability to probe electronic structure with
sub-Angstrom resolution results from the sensitivity of tunnel
current to tip-sample separation and local work function. The
first STM reports of molecule-surface interactions were of
localized chemical reactions with surfaces \cite{Baro,Wolkow99}. 
On Si(111), charge distributed within the $7\times7$ unit cell
both modulates and responds to reaction with ammonia
\cite{Wolkow88}. On the unpinned n-type GaAs (110) surface,
chemisorbed oxygen (being electronegative) images with increased
filled-state density from transferred surface charge and induces
localized surface band-bending \cite{Stroscio}. Patterning of
surface contrast by NH$_3$ dipole fields on GaAs has also been
reported
\cite{Brown}. Spin flip sensitivity of adsorbates to local surface
environment \cite{Heinrich}, and the effect of intermolecular
interactions on surface diffusion
\cite{Mitsui} have been resolved. Underlying two dimensional electron gases
\cite{Repp},  substrate  strain
\cite{Wolkow95,Thayer} and substrate charge transfer 
\cite{Fernandez-Torrente} have 
been found to affect adsorbate pair separation.

Observations of discrete intermolecular interactions and their effect
on STM imaging contrast are limited.  In cryogenic STM work, distance dependent interactions between single Au atoms were studied on NiAl\cite{Niliusa}.  One dimensional (1D) particle in a box states in Au chains on NiAl\cite{Wallis}, and perturbation of these by physisorbed organic molecules have been reported\cite{Niliusb,Nazin}.
Studies of charge transfer 
complexes\cite{Jackel} and coupling 
between functional groups tethered
to molecules are more recent\cite{Lewis}.  STM transport in adjacent silicon atoms was found
to be perturbed by dipole fields due to molecules located
elsewhere in the Si7$\times$7 cell\cite{Harikumar}, and dipole
driven ferroelectric assembly of styrene at 7K has been reported
\cite{Baber}. 

We present experimental (300 K) and theoretical results that show
dipole fields produced by substituents bound to aromatic rings on
styrene molecules significantly perturb transport characteristics
of the host molecules, nearby molecules, and the substrate to
which the molecules are attached.  For one-dimensional
para-substituted OCH$_3$-styrene/CF$_3$-styrene molecular
heterostructures, transport through molecules at the
heterojunction deviates from that of molecules elsewhere within
the structure.  In the case of lines of CF$_3$-styrene arranged
{\em side by side}, the dipole fields increase the ionization potential
of underlying silicon valence electrons.

Organic molecular heterostructures were grown using a vacuum
phase self-directed growth mechanism \cite{Lopinski} for 
styrene on H-terminated \cite{Boland} Si(100)
$2\times1$ surfaces\cite{doped}. Dangling bonds on the H:Si(100)
surface initiate a chain reaction between surface Si atoms
and styrene, leading to well ordered 1D molecular arrays
along Si dimer rows.  Heterostructures were formed by first
dosing CF$_3$ (electron withdrawing) and then OCH$_3$
(electron donating) para-substituted styrene molecules.  Such
substituents are of interest as they modify the energy and
spatial distribution of $\pi$ and $\pi$* states in host aromatic
molecules.

%
\begin{figure}[b]
\includegraphics[width=0.76\linewidth, clip=true, trim=0.0 67 0.0 0.0]{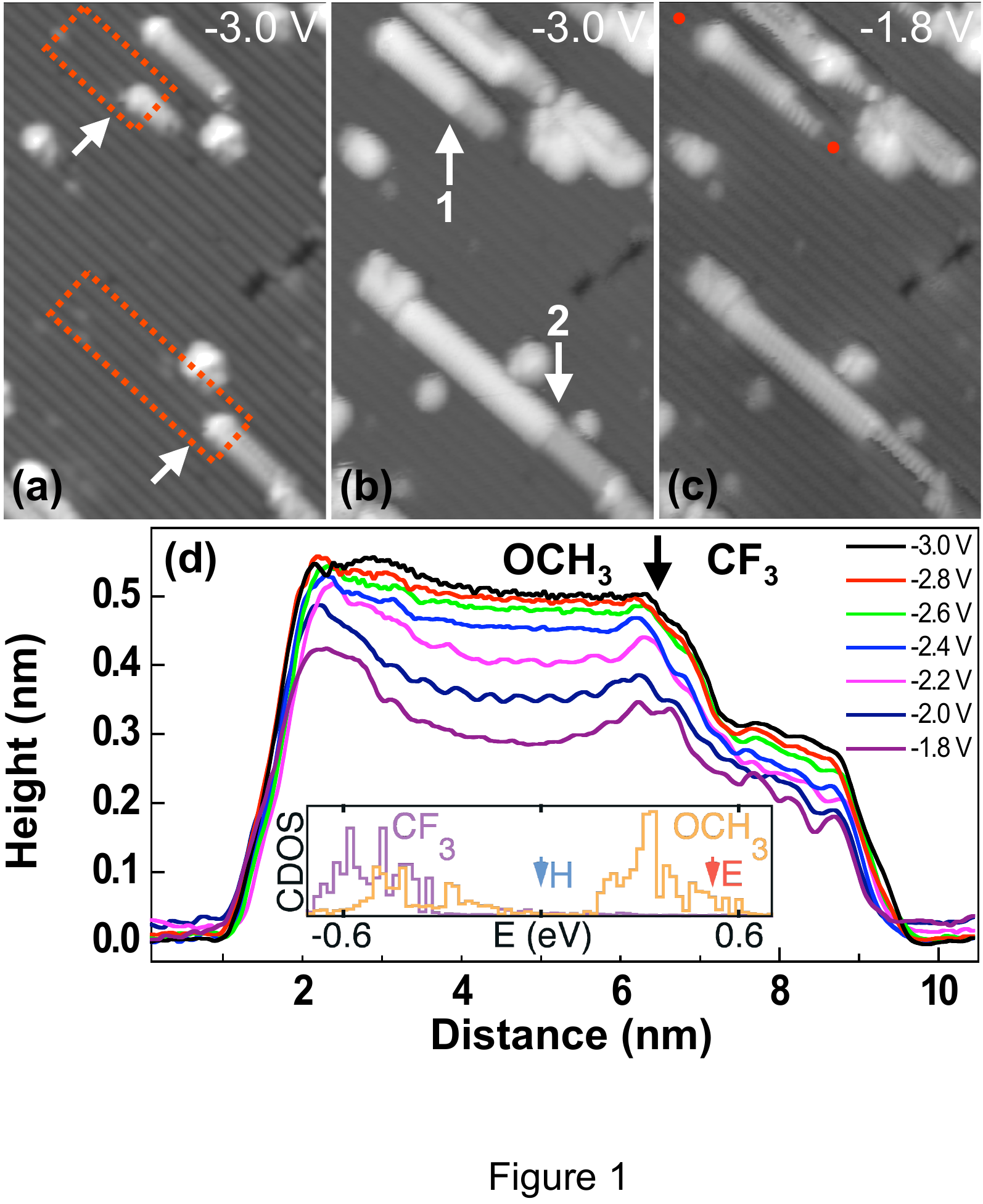}
\caption{\label{Fig_1}Constant current filled-state STM images of CF$_3$-styrene/OCH$_3$-styrene
heterowires on H:Si(100).  (a) Lines of CF$_3$-styrene.  Arrows
indicate reactive dangling bonds.  (b) OCH$_3$-styrene lines have grown in
the red rectangles, extending  the CF$_3$-styrene lines in (a)
to form
CF$_3$-styrene/OCH$_3$-styrene heterowires. (c): As in (b)
imaged at lower bias.  Molecules are bound to right side of
Si dimer row marked by red dots.  (d) Constant-current
topographic cross-sections (0.4nm wide) of heterowire 1 along 
trench to  right of attachment dimers. At low bias, interfacial 
OCH$_3$-styrene (black arrow) images with increased height. Tunnel
current: 40pA. Inset: Molecular HOMO densities of states 
(C orbital projection) of OCH$_3$-styrene (orange) and CF$_3$-styrene (violet) on H:Si.
Arrows H,E show STM tip $E_F$ for plots H,E of Fig. \ref{Fig_2}.}
\end{figure}

Fig.\ref{Fig_1} shows the growth and bias-dependent STM imaging \cite{STM}
of two CF$_3$-styrene/OCH$_3$-styrene heterowires on
H:silicon.  Fig.\ref{Fig_1}(a) shows a 16nm$\times$26nm region
of the sample after a 10L (1L = $10^{-6}$ Torr sec) exposure of
CF$_3$-styrene.  Sample bias $V_s$ was -3.0V.  Arrows
label the reactive dangling bonds at the ends of two
CF$_3$-styrene segments where their growth terminated.  Due to
slight tip asymmetry, CF$_3$-styrene bound to either side of
their host dimers image with slightly different corrugation. 
Comparison with images of the unreacted H:Si surface (not
shown) shows the upper (lower) CF$_3$-styrene segments are
chemically bound to the right (left) sides of
their respective dimer rows.  Fig.\ref{Fig_1}(b) shows the same
region ($V_s$=-3.0V) following a subsequent 10L exposure of
OCH$_3$-styrene. Lines of OCH$_3$-styrene molecules have grown in
regions marked by red rectangles, beginning at the locations
of the terminal dangling bonds of the CF$_3$-styrene lines in
Fig.\ref{Fig_1}(a). Thus two CF$_3$-styrene/OCH$_3$-styrene
heterowires (`1' and `2') have been formed.  

At -3.0 V, the tip Fermi-level is below the highest occupied
molecular orbitals (HOMO) for the OCH$_3$-styrene 
since at this bias the tip height at constant current has saturated (see Fig.\ref{Fig_1}(d)). 
As in our model energy level structure  (inset, Fig.\ref{Fig_1}(d)) 
at high bias the tip Fermi-level (arrow H) 
is below the highest OCH$_3$-styrene HOMO band (orange)
but above the HOMO band of CF$_3$-styrene (violet) which therefore images lower (less bright).

Fig.\ref{Fig_1}(c) shows the same region at $V_s$ = -1.8 V.  Here the tip Fermi-level 
(arrow E, Fig.\ref{Fig_1}(d)) is near the top of the OCH$_3$-styrene HOMO band.  The OCH$_3$-styrene continues to image above
(brighter than) the CF$_3$-styrene, but the OCH$_3$-styrene molecules near the
heterojunctions in heterowires 1 and 2 now image higher than those further away.  The OCH$_3$-styrene in heterowire 1 near the terminal dangling bond also images with increased height.

Fig.\ref{Fig_1}(d) presents topographic cross-sections 
along heterowire 1 above the trench between its attachment row
(labelled with red dots in Fig.\ref{Fig_1}(c)) and the
vacant dimer row to its right.  The topographic envelope for the
heterostructure extends between $\sim$1nm and $\sim$9.5nm along
the abscissa.  The maxima associated with the terminal dangling
bond and the heterojunction are  at $\sim$2.3nm and
$\sim$6.4nm, respectively.  The sloping bias-dependent height response of the
OCH$_3$-styrene segment near the terminal dangling bond is much
like that reported in Ref. \onlinecite{Nature} for styrene:  On
the degenerately doped n-type H:Si surface, dangling bonds
behave as acceptors, and carry negative charge. Molecular
orbitals belonging to molecules in the vicinity of these charge
centres are raised in energy by the localised electrostatic
field.  Therefore at low filled-state bias, these molecules
present increased state density at the tip Fermi-level and image
with increased height.

The bias-dependent height response of the OCH$_3$-styrene near the
heterojunction is similar to that near the terminal dangling bond just
described:  At high bias, the interfacial OCH$_3$-styrene images with
nearly constant height along the bulk of the homowire segment.  As
$|V_s|$ decreases, the height of the interfacial OCH$_3$-styrene (4-5
molecules closest to the heterojunction) does not decay as rapidly as
in the rest of the OCH$_3$-styrene segment.  At $V_s$= -1.8V the
interfacial OCH$_3$-styrene molecules image
$\sim$0.05nm higher than OCH$_3$-styrene situated 5-7 dimers away  from the
heterojunction \cite{Heterowires}. This behavior was not expected as there
is no dangling bond near the junction.

STM imaging characteristics of related 1D styrene/4-methylstyrene
heterostructures on (100) Si were reported in Ref. \onlinecite{stymeth}.  No
height enhancement at the heterojunction was evident in that work. The
perturbation due to the methyl substituent gives rise to weaker
electric dipoles than those investigated here \cite{Anagaw}. This suggests
that the height enhancement at the CF$_3$-styrene/OCH$_3$-styrene junction
may be due to molecular dipoles.

\begin{figure}[b]
\includegraphics[height=0.88\linewidth,angle=90,clip=true, trim=26 0.0 0.0 0.0]{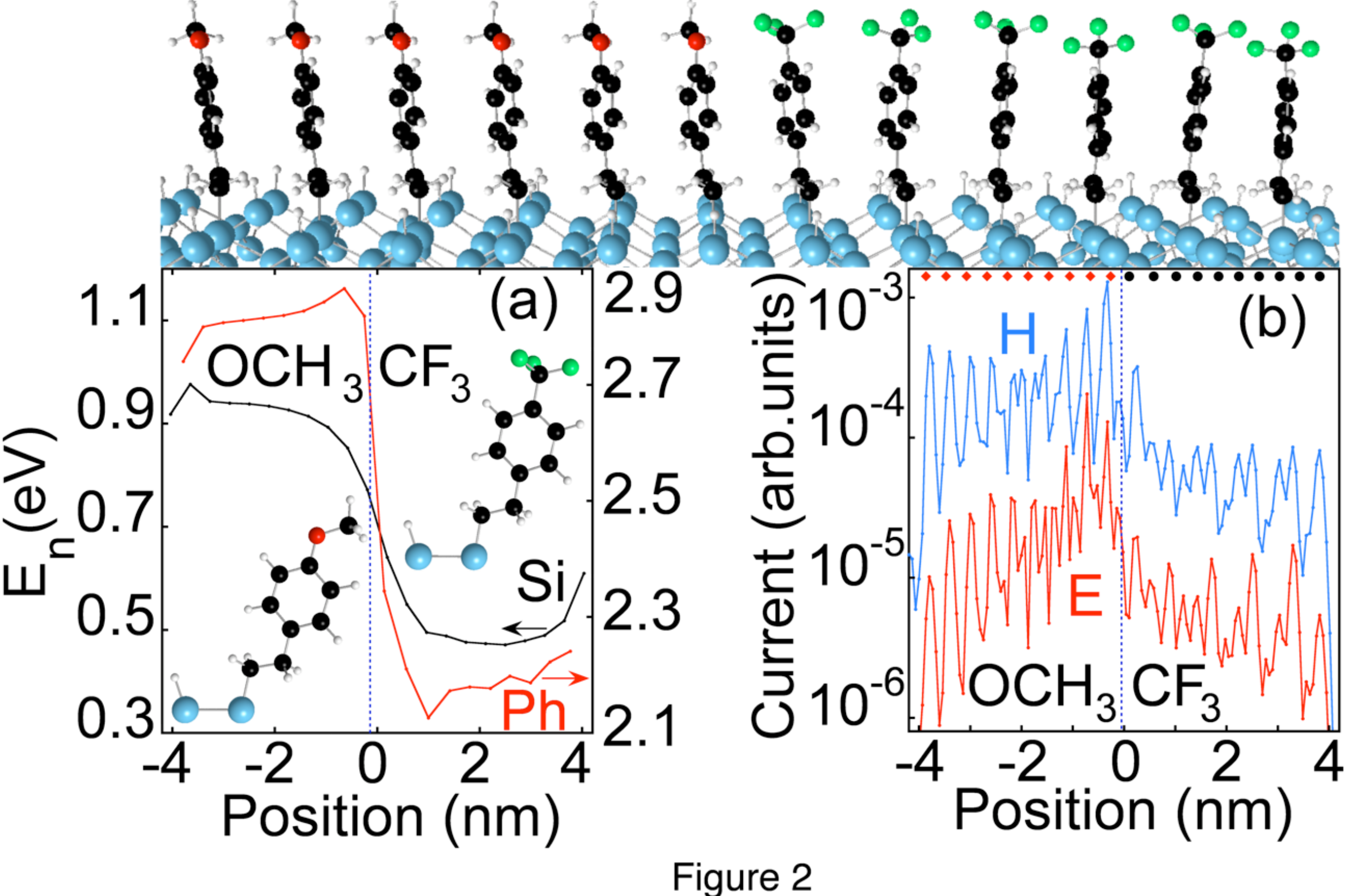}
\caption{\label{Fig_2}(a) Calculated local electrostatic electronic energy shifts $E_n$ vs. position in 
molecular chain. Red curve (Ph): Average of $E_n$ over the benzene ring of each molecule (right scale). Black curve (Si): 
$E_n$ on Si atoms to which molecules bond (left scale). (b) Calculated STM current
$I$ at low (E) and high (H) negative substrate bias vs. STM tip position along molecular chain at constant tip height. Black
bullets (red diamonds) show positions of C (O) atoms of CF$_3$ (OCH$_3$) groups. {\em Part} of the heterostructure near
the junction and side views of CF$_3$-styrene and OCH$_3$-styrene molecules are shown. Si,C,O,F and H atoms are blue, black, red, green and white.}
\end{figure}

To explore this possibility we carried out {\em ab initio} density functional
calculations
\cite{Gaussian} of the electrostatic shifts $E_n = -e(W_n -
U_n)$  of the local electronic energies where $W_n$ ($U_n$) is the electric
potential at the nucleus of atom
$n$  in the presence (absence) of all other atoms of the heterostructure. The results for a 
chain of 10 CF$_3$-styrene and 10 OCH$_3$-styrene molecules on a (100)
H:Si cluster are shown in Fig.\ref{Fig_2}(a); the relaxed geometry \cite{Gaussian} of
a {\em part} of the heterostructure near the junction is shown at the top of
Fig.\ref{Fig_2}.  The red curve in Fig.\ref{Fig_2}(a) shows
$E_n$ averaged over the aromatic ring of each molecule where most of 
the molecular HOMO resides; the OCH$_3$-styrene (CF$_3$-styrene)
molecules are to the left (right) of the blue dotted line in Fig.\ref{Fig_2}(a). 
For sterically favored orientations of the OCH$_3$ dipoles  (negative
O nearer the heterojunction than positive CH$_3$) the red curve rises as the junction
is approached from the OCH$_3$-styrene side, 
peaking  at the 2nd OCH$_3$-styrene molecule from the junction. Hence
the HOMO level of this molecule is higher  in energy than for any other molecule
in the chain. Thus for filled state imaging, as the bias voltage $|V_s|$ increases
the STM tip's Fermi level should cross the HOMO levels of the  OCH$_3$-styrene
molecules near the heterojunction first, resulting in a pronounced low bias peak in
the STM  profile of the heterostructure near the heterojunction on its
OCH$_3$-styrene side, as is seen experimentally in Fig.\ref{Fig_1}(d). 

This is supported by detailed transport simulations: Extended H\"{u}ckel
theory tailored as in Ref. \onlinecite{stymeth} to  describe
the band structures of Si and tungsten and the electronic structures of
molecules, but modified to include the {\em ab initio} electrostatic energy shifts
$E_n$ \cite{Gaussian}, was used to model the
electronic structure of the system. The STM current was then calculated as in Ref.
\onlinecite{stymeth}  solving the Lippmann-Schwinger equation to
determine the electron transmission probability $T(E,V_s)$ between STM tip and
substrate at energy $E$ and bias 
$V_s$. The Landauer expression $I(V_s) = \frac{2e}{h} \int_{-\infty}^{+\infty}
dE\:T(E,V_s)\left( f(E,\mu_{s}) - f(E,\mu_{d})\right)$, where $f(E,\mu_{s})$ and
$f(E,\mu_{d})$ are the source and drain  Fermi functions, was then used to evaluate
the current $I$.     

Fig.\ref{Fig_2}(b) shows the calculated STM current at constant height above the heterowire. Curve E 
is for a tip Fermi-level $E_F$ just below the highest molecular HOMO state (low STM
bias); see the inset, Fig.\ref{Fig_1}(d). As discussed above, resonant transmission via this state (centred on the 2nd
OCH$_3$-styrene molecule from the junction) results in enhanced current there,
consistent with experiment. Curve H  shows the calculated current (at
higher bias) with the tip Fermi-level below the OCH$_3$-styrene HOMO band; see the inset, Fig.1(d). Here
resonant tunneling occurs via the HOMO of {\em every} OCH$_3$-styrene molecule in
the array. Thus the relative interfacial current enhancement decreases as absolute
current levels rise along the chain, again as in the experiment. These effects
result from dipole fields due to OCH$_3$ substituents on the styrene
molecules: Simulations with atomic positions unchanged 
but without electrostatic corrections remove the interfacial
current enhancement entirely.  Simulations with matrix elements and basis function
overlaps responsible for electronic hopping between molecules set to zero did not
significantly modify the results reported here.  This indicates an
electrostatic origin for height enhancement at the heterojunction.
Calculations with styrene replacing the CF$_3$-styrene molecules confirm that 
the interfacial feature is mainly due to 
OCH$_3$-styrene (rather than CF$_3$-styrene) dipole fields \cite{long}.

\begin{figure}[b]
\includegraphics[width=0.67\linewidth,clip=true, trim=0.0 110 0.0 0.0]{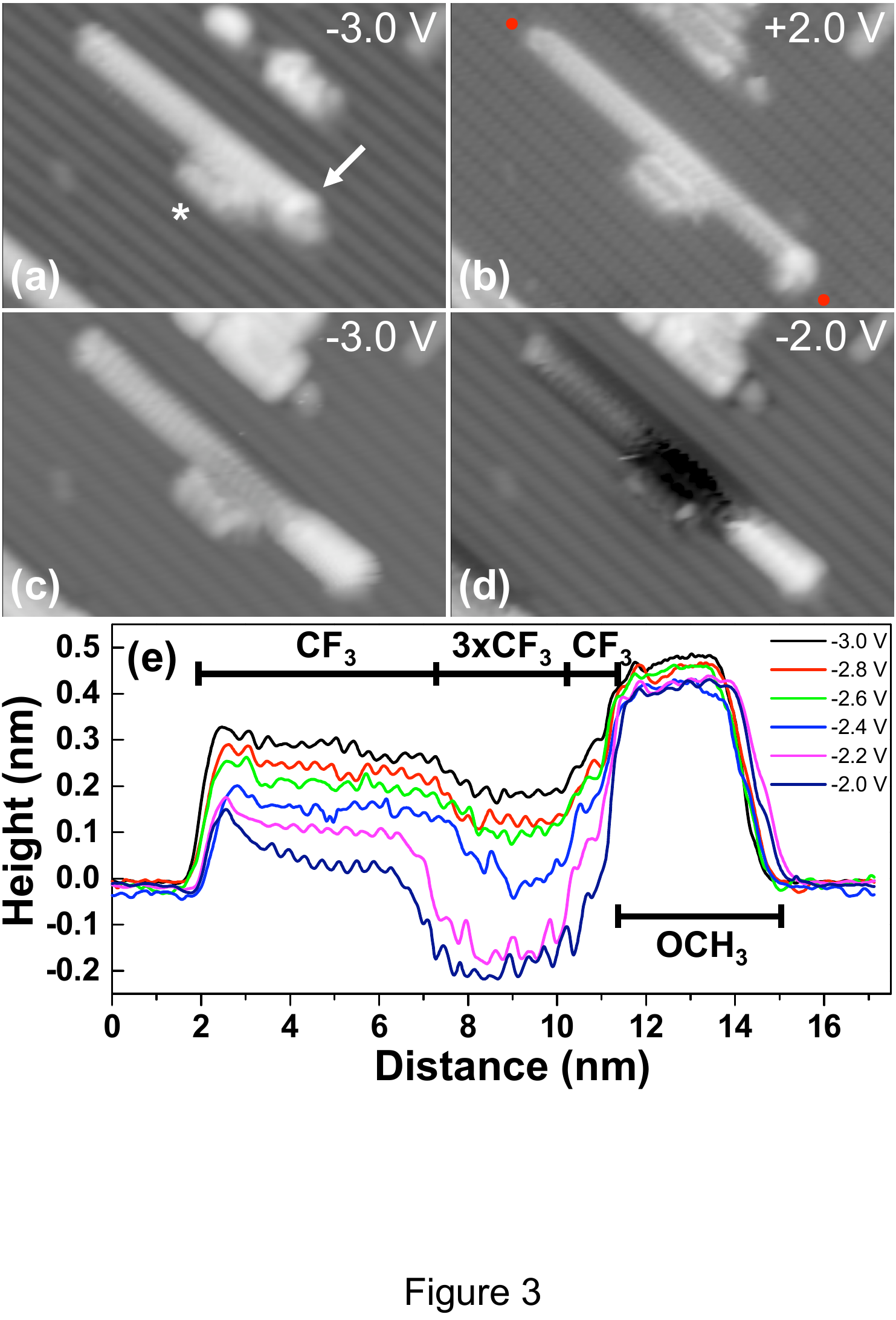}
\caption{\label{Fig_3}STM images of a
(single-triple CF$_3$-styrene)/OCH$_3$-styrene
heterostructure.   (a)
Short double CF$_3$-styrene
line ($\star$) beside longer single CF$_3$-styrene chain.
Arrow marks dangling bond.   
(b)$V_s$=+2V. Long CF$_3$-styrene chain 
extended by $\sim$7 OCH$_3$-styrene molecules.  
(c) $V_s$=-3V. OCH$_3$-styrene images above CF$_3$-styrene.  Single
and triple CF$_3$-styrene lines image with similar
height.  (d) $V_s$=-2V.  Single 
OCH$_3$-styrene and CF$_3$-styrene lines still image above
H:Si surface (brighter).  Triple CF$_3$-styrene chains
 image below H:Si surface (black). 
(e) Constant current topographic cross-sections (0.4nm
wide) of 
CF$_3$-styrene/OCH$_3$-styrene heterowire along trench
to right of attachment dimers.  Heights are 
relative to H:Si surface (height = 0nm). Tunnel current: 40pA.}
\end{figure}

Effects of the dipole fields are not limited to molecular energy levels:  The
black curve in Fig.\ref{Fig_2}(a) suggests that the Si valence states 
may lie
$\sim$0.5 eV lower under the CF$_3$-styrene chain than under the
OCH$_3$-styrene.  A related heterostructure studied below highlights the
response of the underlying silicon to the molecular dipole fields: Fig.\ref{Fig_3}
shows STM imaging of a triple/single chain of CF$_3$-styrene molecules  on
H:Si(100).  Fig.\ref{Fig_3}(a) shows a 15nm$\times$10nm region 
following a 10L exposure of CF$_3$-styrene ($V_s$= -3.0V).  The arrow points
to the reactive dangling bond at the end of the longest CF$_3$-styrene
line.  The $\star$ marks a short double chain of
CF$_3$-styrene that has grown beside the long CF$_3$-styrene chain. 
Figs.\ref{Fig_3}(b)-(d) show the same region following a 10L exposure of
OCH$_3$-styrene.  The end of the long CF$_3$-styrene chain has been extended by
$\sim$7 molecules of OCH$_3$-styrene.  Figs.\ref{Fig_3}(b) and (c) ($V_s$= +2V and
-3V, resp.) image the single and triple CF$_3$-styrene segments with comparable
height.  In Fig.\ref{Fig_3}(d), $V_s$ has been reduced to -2.0V and the region with
the triple CF$_3$-styrene lines images below (darker than) the single file chain of
CF$_3$-styrene.  Fig.\ref{Fig_3}(e) shows topographic cross-sections along the
CF$_3$-styrene/OCH$_3$-styrene heterowire.  From $V_s$=-3V to $V_s$=-2V, the
triple CF$_3$-styrene chain (between 7nm and 10nm) images with decreasing
height.  At $V_s$ = -2.0V, this region images 0.2nm below the H:Si surface 
indicating depleted silicon state density beneath the molecules at the tip
Fermi level.

\begin{figure}[t]
\includegraphics[height=0.94\linewidth, angle=90,clip=true, trim=75 0.0 0.0 0.0]{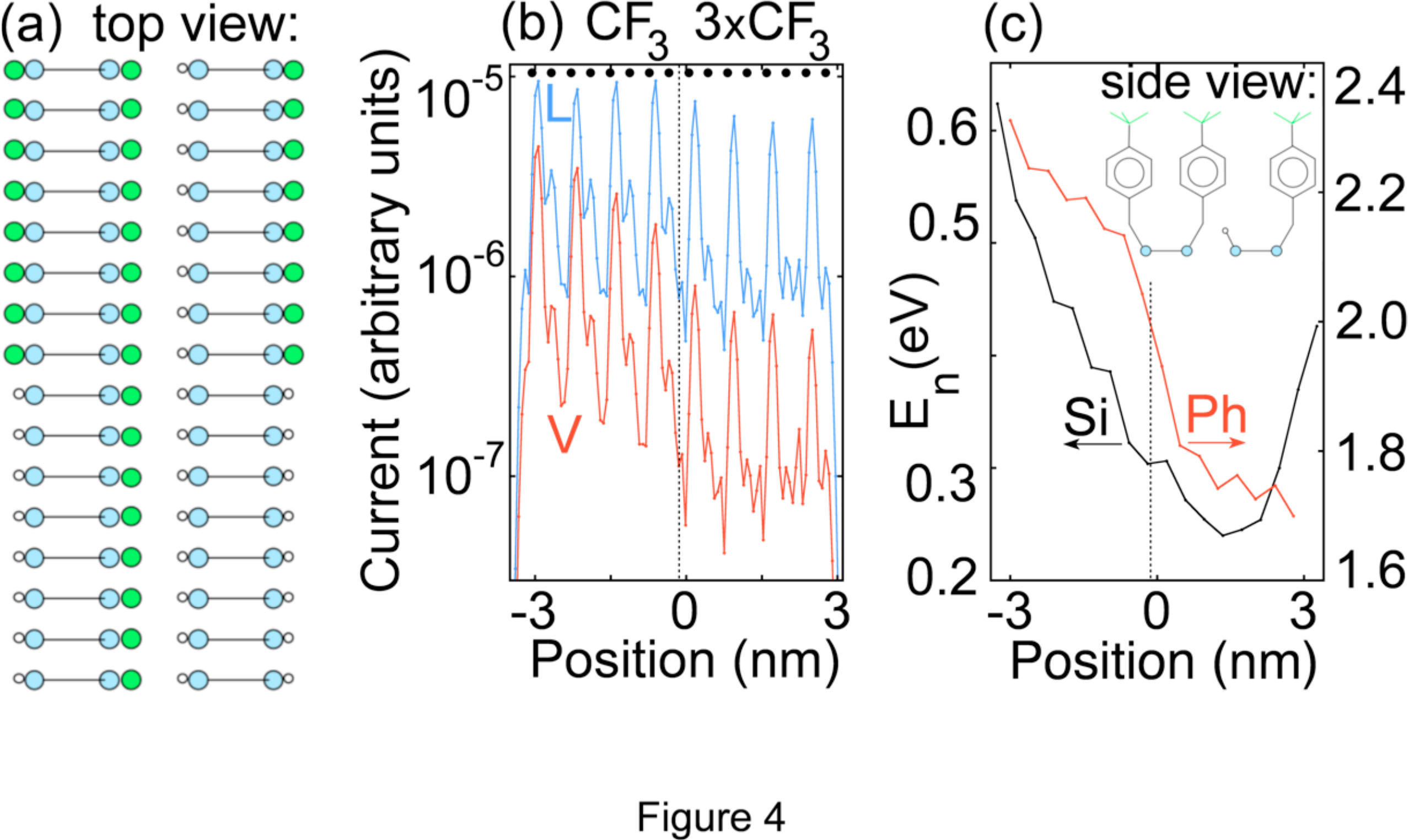}
\caption{\label{Fig_4}(a) and inset: Schematic of 
model single-triple CF$_3$-styrene structure. (b) Calculated current profiles. 
Plot V: Low bias; tip Fermi level near highest Si valence
band states. Plot L: Higher bias; tip Fermi level above
CF$_3$-styrene HOMO energies.  As in experiment, contrast between triple and
single CF$_3$-styrene rows in blue  profile is much weaker than in red
profile.  Black bullets locate C atoms of CF$_3$ groups.  (c) Black
curve: Electrostatic electronic energy shifts $E_n$ at Si atoms to
which molecules of long CF$_3$-styrene row bond. Triple (single)
rows are right (left) of dotted line. Red curve as in Fig.\ref{Fig_2}.}
\end{figure}

Transport simulations were undertaken for the related structure shown in Fig.\ref{Fig_4}(a) (the long
CF$_3$-styrene line is between the short ones to minimise
sensitivity to cluster edges).  Fig.\ref{Fig_4}(b) shows the simulated constant height current 
along the long CF$_3$-styrene line. At low bias (curve V), current levels drop over
the triple CF$_3$-styrene qualitatively as in  Fig.\ref{Fig_3}.  The origin of this is seen in Fig.\ref{Fig_4}(c):  Dipole fields
of the CF$_3$-styrene lower Si orbital energies below the triple
CF$_3$-styrene by
$\sim$0.2eV more than under the single file CF$_3$-styrene. 
At low bias ($V_s$$\sim$-2.0V) this reduced silicon state density at the tip
Fermi level forces the STM tip (in experiment) to move lower over triple CF$_3$-styrene than over 
the H:Si surface to re-establish the fixed tunnel current (40 pA).  It can also be concluded
in this regime that lateral carrier transfer from the single chain CF$_3$-styrene and
OCH$_3$-styrene regions to the triple chain CF$_3$-styrene region is negligible
compared with the direct through-molecule transport component.

In summary, we have shown experimentally and theoretically that dipole fields established by strongly electron donating or withdrawing chemical species bound to molecules significantly modulate
the transport characteristics of nearby molecules and the underlying substrate.  Judicious selection of substituents
attached in a site specific manner can be used to tailor electron transport at the molecular length scale and allows differential tuning of
molecular vs. substrate transport characteristics.  

This research was supported by CIFAR, NSERC, iCORE, Westgrid and the NRC. We
have benefited from discussions with G. DiLabio and from the  technical expertise of D. J. Moffatt and M. Cloutier.

%
%

%

\begin{thebibliography}{100}
\bibitem{doped}The Si was 
arsenic-doped with resistivity $< 0.005
\Omega$ cm.
\bibitem{STM}STM imaging was in vacuum 
$<$1$\times$$10^{-10}$Torr with W tips.
\bibitem{Heterowires}Heterowires were studied on multiple 
H:silicon surfaces with different STM tips. 
Additional images will be presented elsewhere\cite{long}.
\bibitem{Gaussian} Gaussian03 with the B3PW91 functional
and Lanl2DZ basis was used for electrostatic calculations.
Gaussian98 with the UFF model was used to relax geometries.
\end{thebibliography}
\end{document}